\documentclass{aa}
\usepackage{graphics}
\begin{document}

   \thesaurus{05     
              (08.04.1;  
               10.15.1;)} 
   \title{Open clusters with Hipparcos
   \thanks{Based on observations made with the 
   ESA Hipparcos astrometry satellite}}

   \subtitle{I. Mean astrometric parameters}

   \author{N. Robichon \inst{1,2} \and F. Arenou \inst{2} 
   \and J.-C. Mermilliod \inst{3} \and C. Turon \inst{2} }

   \offprints{No\"el Robichon}

   \institute{
          Sterrewacht Leiden, Postbus 9513, NL-2300 RA Leiden, The Netherlands
        \and
   	Observatoire de Paris, section de Meudon, DASGAL/CNRS URA 335, 
		F-92195 Meudon CEDEX, France\\
		email Noel.Robichon@obspm.fr, Frederic.Arenou@obspm.fr, 
		Catherine.Turon@obspm.fr
         \and
		Institut d'Astronomie de Lausanne, CH-1290 Chavannes des Bois, 
		Switzerland\\
		email Jean-Claude.Mermilliod@obs.unige.ch}

   \date{Received ; accepted }

   \maketitle

   \begin{abstract}

New memberships, mean parallaxes and pro\-per motions of all 9 open clusters
closer than 300~pc (except the \object{Hyades}) and 9 rich clusters between
300 and 500~pc have been computed using Hipparcos data. Precisions,
ranging from 0.2 to 0.5~mas for parallaxes and 0.1 to 0.5~mas/yr for
proper motions, are of great interest for calibrating photometric
parallaxes as well as for kinematical studies.
Careful investigations of possible biases have been performed
and no evidence of significant systematic errors on the mean
cluster parallaxes has been found.
The distances and proper motions of 32 more distant clusters, 
which may be used statistically, are also indicated.
   \end{abstract}

%
\section{Introduction}
%
\def\lynga{Lyng{\aa} }

Hipparcos observations of stars in nearby open clusters offer, for the 
first time, the possibility of determining accurate distances to these clusters
without any assumption about their chemical composition or about stellar 
structure. The new distance modulus of the \object{Hyades}, 3.33 $\pm$ 0.01, derived
by Perryman et al (\cite{perryman}) is a first step in the determination of the distance 
scale in the universe. The high precision obtained represents an important
improvement with respect to the results of decades of attempts to fix the 
zero point of the distance scale.

The position of the Zero Age Main Sequence (ZAMS) is sensitive to the
exact chemical composition
of the clusters and a difference of [Fe/H] = 0.15,
corresponding to the metallicity difference between the \object{Hyades} and the Sun, 
results in a displacement of about 0.2 magnitude in absolute
magnitude $(M_V)$ according to several internal structure and
atmosphere models. As the exact chemical composition of most clusters is
not presently known with the required accuracy, the metallicity
corrections to the distance moduli are not known with precision.
Thanks to Hipparcos observations, it is possible to determine the
absolute position of the main sequences of several open clusters
independently of any preliminary knowledge of the chemical composition.
According to the present data on chemical composition, no large
discrepancies are found between the Hipparcos distance moduli of most of
the cluster and the positions of their sequences in the HR diagram
(Mermilliod et al. \cite{mermilliod97a}, Robichon et al.
\cite{robichon97}), with the noticeable exception of the Pleiades.
Because the Main-Sequence Fitting (MSF) method is still
the basic tool in determining the 
distances of open clusters, the understanding of the Pleiades anomaly 
appears to be the first priority.

Pinsonneault et al. (\cite{pinsonneault}) (herafter PSSKH) have tackled
the problem with a grid of models adapted to the mass range of solar-type
stars which are unevolved in nearby clusters, and chemical
composition of these clusters. Their method determines the distance modulus
and metallicity simultaneously from $(M_V, (B-V)_0)$ and $(M_V, (V-I)_0)$,
using the fact that $(V-I)$ is much less sensitive to the metallicity
than $(B-V)$. Good agreement is found
for several clusters (\object{Hyades}, \object{Praesepe}, \object{$\alpha$ Per}sei), i.e. the
distances determined for the adopted metallicity correspond to those
obtained from Hipparcos. Problems are found for the \object{Pleiades} (and Coma
Ber cluster which only has $B-V$ colours). PSSKH attributed these discrepancies
to 1 mas systematic errors in the Hipparcos Catalogue.

In fact, a more general view of
the situation should be obtained from the analysis of additional nearby
open clusters. For example, \object{NGC 2516} which occupies the
lowest position in the HR diagram with respect to \object{Praesepe} (even below
that of the \object{Pleiades}) has a metallicity $[Fe/H] = -0.32$
(Jeffries et al. \cite{jeffries}), in good agreement with that required to
adequately fit the ZAMS in the colour-magnitude diagram.

The results and detailed discussions presented in this paper are in
keeping with preliminary results presented at the Venice'97 Symposium
(Robichon et al. \cite{robichon97}). Since this Symposium, careful investigations of possible
biases have been performed, but no evidence of any bias larger than few
tenths of a milliarcsecond has been discovered. Discrepancies between
the parallaxes of the \object{Pleiades} and \object{Coma Ber} with the ground-based values
of Pinsonneault et al. still exists, and an attempt to explain them will
be given in a following paper (Robichon et al. in prep.).
This second paper will analyse the cluster
sequences in the colour-magnitude diagram in the light of Hipparcos data.
It will complete the analysis of the cluster sequences in several photometric
systems presented in Mermilliod (\cite{mermilliod98}) which  
exhibits a significant correlation between the cluster metallicities and their
relative positions in the $(M_V,~{(B-V)_0)}$ diagram when using the Hipparcos
distance moduli.

The outline of the paper is the following. Section \ref{selection} depicts the two
different methods adopted for selecting cluster members from the Hipparcos astrometric
data, depending on whether or not they are closer than 500 pc and contain
at least 8 members.
With these sets of members, the mean astrometric parameters 
($\pi$, $\mu_\alpha\cos\delta$, $\mu_\delta$) of 18 rich clusters
closer than 500 parsecs, and 32 more distant and/or containing between 4 and 7 members,
are computed and given in Sect. \ref{param}. The method used to compute these
mean astrometric parameters is briefly described. It utilizes Hipparcos
intermediate data which allow to take account of the star to star correlations.
The rest of the paper reviews the possibility of systematic errors in the parameters both at
large scale and small scale. The conclusion of this last
part is that the mean astrometric parameters are statistically unbiased over the sky and
that their formal errors are not severely underestimated.

%
\section{Selection of cluster members}\label{selection}
%
%
\subsection{Pre-launch selection}
%
The initial selection for inclusion of cluster stars in the Hipparcos
Input Catalogue (HIC) (Turon et al. \cite{turon}) is described in detail
in (Mermilliod \& Turon \cite{mermilliod89}).
It was based on the conditions of membership from
proper motions and radial velocities when available, and the positions in
the colour-magnitude diagram on the single star sequence to minimize the
effects of potential companions. Further selections were applied during the
mission simulations to remove those stars that could be affected by
veiling glare of bright neighbouring stars. In the case of the \object{Pleiades}
and \object{Praesepe}, stars from the outer region have been included in the sample
to enlarge the total number of stars in these two clusters.
As in any other field, the selection was also constrained by the
satellite capabilities and was achieved through simulations.

The candidates in the \object{Praesepe} and \object{Pleiades} clusters were selected on
the basis of 
proper motion, radial velocity and photometry analysed in Mermilliod 
et al. (\cite{mermilliod90}) and Rosvick et al. (\cite{rosvick}),
with the same criteria, especially concerning the duplicity. These
conditions are reflected in the fact that the sequences in the
colour-magnitude diagrams of most clusters are quite narrow.

%
\subsection{Final catalogue member selection}
%
In this study, two different member selections have been applied
to  the open clusters in order to securely
distinguish the members from the field stars based on their astrometric
parameters ($\pi$, $\mu_\alpha\cos\delta$, $\mu_\delta$).

The mean astrometric parameters of clusters closer than 500 pc and containing 
at least 8 stars observed by Hipparcos can be derived with good
accuracy. Because they are quite different from field star parallaxes and
pro-per motions, a new and secure selection of members in the Hipparcos
Catalogue can be performed, which replaces the pre-launch selected sample.
This concerns all the clusters closer than 300 pc and 8 additional clusters
closer than 500 pc.

For the other clusters, situated further than 500 pc or with
a number of Hipparcos stars smaller than 8, the mean parallaxes and proper
motions are small or not accurate enough and members are harder to separate
from field stars on an astrometric basis. A selection based only on
astrometrical criteria would accept non member stars and could then bias
the computed mean parameters of the cluster. Nevertheless, even if the
mean Hipparcos parallax is not so precise compared to distance modulus derived,
for example, from a MSF, it is interesting to compute their
mean astrometric parameters for at least two reasons. On the one hand,
mean parallaxes of dozens of clusters allow statistical calibration of
other distance indicators. On the other hand, the cluster mean proper
motions can be very useful for galactic kinematic studies. For these
clusters, only stars preselected in the Hipparcos Input Catalogue were
taken into account. For the 110 clusters farther than 300 pc and 
with at least 2 Hipparcos stars, the mean astrometric parameters
have also been derived. 

No attempt has been made to find new nearby clusters in the Hipparcos
Catalogue. Platais et al. (\cite{platais98}) made a survey of new
open clusters and associations in the Hipparcos Catalogue. They found
some possible new clusters which need to be confirmed by further
analysis at fainter stars. These new objects are then not included in the 
present paper. The same goes for OB associations which are studied
in detail using Hipparcos data in a comprehensive paper by de Zeeuw et al.
\cite{dezeeuw}. The method used here to derive cluster mean astrometric parameters
is not suited for the \object{Hyades} because its depth is not a negligible 
fraction of its distance at the Hipparcos precision. The \object{Hyades} 
properties were analysed in detail by Perryman et al. (\cite{perryman})
with the Hipparcos data.

The selections carried out in this paper rely on the assumption that
all the cluster members have the same space velocity and, for the 
closest clusters, that they lie within a 10 parsec radius sphere centred on the
cluster centre (which roughly corresponds to the tidal radius of an open
cluster). One cluster, \object{NGC 1977}, has been rejected from the present study
because the distribution of its
members over the sky is not in good agreement with a bound cluster
(in particular, no centre can be defined). These stars are rather part of
a 80 pc long feature, connected with the \object{Orion OB1} association
(Tian et al. \cite{tian}).
Another nearby object, \object{Melotte 227}, as well as most of the nearby
Collinder groups (Cr 399, 359, 135 and 463) have been rejected 
since the astrometric data of the preselected stars do not show the
characteristics of an open cluster, in particular in their spatial structure. 

%
\subsubsection{Members in the closest clusters}
%
Although a visual examination of the vector-point and colour-magnitude diagrams
can easily confirm the presence of an open cluster, an objective selection of
members is always an issue. The selection presented in this section is based
on an iterative method, which converges after 2 or 3 iterations, namely
when no more stars are rejected from the selection. This iterative procedure
is primed with a set of well known members.

\newcommand{\Z}[1]{$^{(#1)}$}

\begin{table}[htbp]
\begin{center}
\caption{Equatorial coordinates (J2000.0) from \lynga (1987)
and mean radial velocity of the cluster centres}
\begin{tabular}{lrrrr}
\hline
Cluster & \multicolumn{1}{c}{$\alpha$} & \multicolumn{1}{c}{$\delta$}
 & \multicolumn{1}{c}{V$_R$} &  \multicolumn{1}{c}{\# of} \\
name & \multicolumn{1}{c}{h m s} & \multicolumn{1}{c}{\degr \arcmin}
 & \multicolumn{1}{c}{km/s} & \multicolumn{1}{c}{stars}\\
\hline
\object{Coma Ber}     & 12 25 07 &  26 06.6 &  -0.1 $\pm$ 0.2 & 22\Z{1}\\
\object{Pleiades}     &  3 47 00 &  24 03.0 &   5.7 $\pm$ 0.5 & 78\Z{1}\\
\object{IC 2391}      &  8 40 14 & -53 03.6 &  14.1 $\pm$ 0.2 & 15\Z{1}\\
\object{IC 2602}      & 10 43 12 & -64 24.0 &  16.2 $\pm$ 0.3 & 18\Z{1}\\
\object{Praesepe}$^*$     &  8 40 00 &  19 30.0 &  34.5 $\pm$ 0.0 & 104\Z{1}\\
\object{NGC 2451}     &  7 45 12 & -37 58.2 &  28.9 $\pm$ 0.7 &  5\Z{2}\\
\object{$\alpha$ Per} &  3 22 02 &  48 36.0 &  -0.2 $\pm$ 0.5 & 18\Z{1}\\
\object{Blanco 1}     &  0 04 24 & -29 56.4 &   5.1 $\pm$ 0.2 & 28\Z{1}\\
\object{NGC 6475}     & 17 53 43 & -34 48.6 & -14.7 $\pm$ 0.2 & 40\Z{1}\\
\object{NGC 7092}     & 21 32 12 &  48 26.4 &  -5.4 $\pm$ 0.4 &  7\Z{1}\\
\object{NGC 2232}     &  6 26 24 &  -4 45.0 &  21.0 $\pm$ 0.6 &  4\Z{2}\\
\object{IC 4756}      & 18 38 58 &  -5 27.0 &  25.8 $\pm$ 0.2 & 13\Z{1}\\
\object{NGC 2516}     &  7 58 00 & -60 48.0 &  22.7 $\pm$ 0.4 &  6\Z{2}\\
\object{Trumpler 10}  &  8 47 48 & -42 29.4 &  25.0 $\pm$ 3.5 &  2\Z{2}\\
\object{NGC 3532}     & 11 06 24 & -58 42.0 &   3.1 $\pm$ 2.5 &  3\Z{2}\\
\object{Collinder 140}&  7 23 55 & -32 12.0 &  19.9 $\pm$ 3.1 &  4\Z{2}\\
\object{NGC 2547}     &  8 10 48 & -49 18.0 &  14.4 $\pm$ 1.2 &  5\Z{2}\\
\object{NGC 2422}     &  7 36 36 & -14 30.0 &  29.4 $\pm$ 3.7 &  4\Z{2}\\
\hline
\label{table1}
\end{tabular}
\end{center}
Source of the mean radial velocity:\\
\Z{1} CORAVEL mean value of members selected from CO\-RA\-VEL and 
photometric data;\\
\Z{2} mean value from the WEB Catalogue (Duflot et al. \cite{web}) of 
selected Hipparcos members.\\
$^*$ coordinates from Raboud \& Mermilliod \cite{raboud}.\\
the last column indicates the number of members used to compute the mean radial velocity.\\
\end{table}

At each iteration, the cluster mean parallax and mean proper motion at
the position of the centre are computed from Hipparcos intermediate
data, according to the computation described in Sect. \ref{param},  using
the members selected at the previous iteration. The computation also
takes into account the cluster mean radial velocity to correct the
perspective effect due to the different angular directions of the
members compared to the cluster centre. The values of cluster centres
are taken from \lynga (\cite{lynga87}) except for the \object{Pleiades} for
which it is taken from Raboud \& Mermilliod (\cite{raboud}) who
derived a new centre of mass for the cluster. These
cluster centres are fixed once for all and are not calculated from
Hipparcos data since the number of Hipparcos stars in each cluster is 
generally not large enough to obtain new accurate
cluster mass centres.

When available, radial velocities obtained with the CO\-RA\-VEL
radial-velocity scanner by Rosvick et al. (\cite{rosvick}), Mermilliod
et al. (\cite{mermilliod97b}) and additional unpublished data have
been used to compute cluster mean radial velocities $V_{R0}$.

Because the CORAVEL scanner is adapted to measure stars later than the
spectral type F5 $(B-V > 0.45)$ while Hipparcos measured the brightest
and thus bluest part of the main sequence,
there are few stars in common between the CORAVEL and HIPPARCOS samples
for most of the clusters.
Therefore, CORAVEL mean velocities have been computed from all observed 
known members (not only Hipparcos members), with the exclusion of binaries
 without a determination of orbital elements. When too few
CORAVEL data were available, radial velocities from the
WEB Catalogue (Duflot et al. \cite{web}) of the Hipparcos stars selected in
this paper were averaged. Fortunately, the value of the mean cluster
radial velocities does not need
to be so precise since only its projection at the position of each
member is used. For example, an error of 1 km/s in the mean radial
velocity would induce an error on the proper motion of a \object{Pleiades}
member situated at 3 degrees from the cluster centre (6 pc) of
about 0.1 mas/yr. Mean radial velocities and the number of stars and
references of the sources (Coravel or the WEB Catalogue) used to compute
them are given in Table \ref{table1}.

Each star in the area of the cluster is submitted to a succession of
selection tests described hereafter and taking into account its
position, parallax, proper motion and photometry, their associated
errors, and the mean radial velocity of the cluster.

Let $\mathbf{x}_i = (\pi_i, {{\mu_\alpha}_i \cos \delta}_i,
{\mu_\delta}_i)$ be the vector containing the Hipparcos parallax and
proper motion of star $i$ with $\mathbf{\Sigma}_i$ being the covariance
matrix.
Let ${\bf x_0} = (\pi_0, {\mu_{\alpha 0} \cos \delta_0}, {\mu_{\delta
0}})$ be the parallax and proper motion of the cluster centre
corresponding to the mean velocity of the cluster with
covariance matrix $\mathbf{\Sigma_0}$. Let $\mathbf{x_0}_i =
({\pi_0}, {\mu_{{\alpha 0}_i} \cos \delta}_i, {\mu_{\delta 0}}_i)$ be
the parallax and proper motion corresponding to the mean velocity of the
cluster at the position of star $i$ with covariance matrix
$\mathbf{\Sigma_0}_i$. These are deduced from $\mathbf{x_0}$ and
$\mathbf{\Sigma_0}$ and the mean radial velocity of the cluster $V_{R0}$ by the
following rotation:

\begin{eqnarray} \mu_{{\alpha 0}_i} \cos\delta_i & = & \cos \Delta \alpha_i 
\mu_{\alpha 0} \cos \delta_0  \nonumber \\ 
&& + \sin\delta_0\sin\Delta\alpha_i \mu_{\delta 0}  \nonumber \\ 
&& -\cos\delta_0\sin\Delta\alpha_i  {V_{R0}\pi_0 \over 4.74} \\ 
\mu_{{\delta_0}_i} & = & -\sin\delta_i\sin\Delta\alpha_i 
\mu_{\alpha_0} \cos \delta_0 \nonumber \\ 
&&+(\cos\delta_i\cos\delta_0+\sin\delta_i\sin\delta_0\cos\Delta\alpha_i)  
\mu_{\delta_0} \nonumber \\ 
&&+(\cos\delta_i\sin\delta_0-\sin\delta_i\cos\delta_0\cos\Delta\alpha_i) {V_{R0}
\pi_0\over 4.74} \nonumber \end{eqnarray}

\noindent where $(\alpha_i, \delta_i)$ are the equatorial coordinates of star $i$,
$(\alpha_0, \delta_0)$ are the equatorial coordinates of the cluster
centre and $\Delta\alpha_i = \alpha_i-\alpha_0$. Note that, since the
cluster depth is neglected, all the members are assumed to share the
same parallax $\pi_0$.

Assuming a Gaussian distribution of errors, the value $\chi^2 =
(\mathbf{ x}_i-\mathbf{x_0}_i)^{\mathbf{T}}(\mathbf{\Sigma}_i +
\mathbf{\Sigma_0}_i)^{-1}(\mathbf{x}_i-\mathbf{x_0}_i)$ follows a
Chi-square distribution with 3 degrees of freedom. Star $i$ is
considered as a
cluster member if $\chi^2 < 14.16$ (corresponding to a 3$\sigma$
Gaussian two-sided test).

If star $i$ is considered as a cluster member at the previous iteration,
i.e. if it is used for the calculation of $\mathbf{\Sigma_0}_i$,
then $\mathbf{\Sigma}_i$ and $\mathbf{\Sigma_0}_i$ are correlated.
Nevertheless this correlation is small and has been neglected because
$\mathbf{\Sigma_0}_i$ is calculated with a sufficiently large number of
stars (between 8 and 54).

A diagonal correlation matrix simulating the depth of the cluster and
the internal velocity dispersion can be added to $\mathbf{\Sigma}_i$ and
$\mathbf{\Sigma_0}_i$ but it is small compared to them. For example,
using 2 pc as a typical cluster core radius and a velocity
dispersion of 0.5 km/s, the member selection in each cluster remains
unchanged.

To avoid any erroneous selection, stars with a standard error of the
parallax larger than 3 mas or a standard error of the proper motion
larger than 3 mas/yr have also been rejected. This concerns only
1 or 2 stars per cluster at the most.

To be sure that only real members are selected, and not stars from a
possible moving group associated with the cluster, stars whose distance
from the cluster centre, perpendicularly to the line of sight, is
greater than 10 pc  (corresponding to a typical open cluster tidal
radius) have also been rejected.

Hipparcos double stars were rejected when their duplicity could damage or
bias the
mean proper-motion and parallax values, i.e. when the field H59 of the
Hipparcos Catalogue was equal to C, G, O, V or X (see ESA \cite{hip}).
G, X and V entries have abscissae on the Reference Great Circles (RGC)
which reflect the combination of the proper motion of the system
(depending of the mean cluster velocity) and the orbital motion of the
system. C and O entries have a proper motion in the Hipparcos main
Catalogue decoupled from the orbital motion because they were reduced
with an appropriate algorithm taking into account more than the 5
astrometric parameters needed to describe the astrometry of a single
star. The global cluster reduction (Sect. \ref{param}) doesn't take these
supplementary parameters into account and thus, the astrometric
parameters that are calculated for these stars could be biased by the
orbital motions. Hipparcos double stars considered as cluster members
and not used in the reduction are given in the appendix Table \ref{table9}.

Once cluster members have been selected from their astrometric parameters,
their membership is verified with the help of the ($V, B-V$) colour-magnitude
diagram (hereafter CMD). This test was positive for all clusters, except for \object{NGC 6475}
in which two stars (HIP 86802 and 88224) were rejected according to their
discrepant positions in the CMD.

The cluster members obtained by this selection process are listed (by HIP
number) in the appendix Table \ref{table10}. The number of selected
members varies from 8 to 54.

%
\subsubsection{Selection of members in more distant clusters}
%
For the most distant clusters or for clusters with less than 8 members, 
the selection is more difficult.
Because it is not possible to redefine a secure membership
selection for these clusters, only the HIC preselected stars were taken into
account and no attempt has been done to identify new members.
Possible non-members were excluded using BDA, the open cluster database 
(Mermilliod \cite{mermilliod95}).
An iterative procedure was then applied to compute the mean
proper motion of the members and to reject stars with a proper motion
discrepant by more than 3$\sigma$ from the mean. The final mean
astrometric parameters are then computed in the same way as for the
nearby clusters (see Sect. \ref{param} below). They may be useful mainly 
for statistical studies (e.g. section~\ref{distant}).
%
\section{Cluster mean astrometric parameters}\label{param}
%
%
\subsection{Hipparcos Intermediate Data}\label{inter}
%
The mean cluster parallax cannot be computed without caution from the
Hipparcos observations. As was
explained before the satellite launch, the estimation of the mean parallax
or proper motion of a cluster observed by Hipparcos must take into account the
observation mode of the satellite (Lindegren \cite{lindegren88}). This is
due to the fact that stars within a small area in the sky have frequently
been observed in the same field of view of the satellite. Consequently, one
may expect correlations between measurements done on stars separated by a few
degrees, or with a separation being a multiple of the basic angle (58\degr)
between the two fields of view.

The consequence is that, when averaging the parallaxes or proper motions
for $n$ stars, the improvement factor does not follow the expected
$1/\sqrt{n}$ law and will not be asymptotically better than
$\sqrt{\rho}$ if $\rho$ is the mean positive correlation between data
(Lindegren \cite{lindegren88}). Ignoring
these correlations would thus underestimate the formal error on the 
average parallax.

The proper way to take these correlations into account is to go one step
back in the Hipparcos reduction and to work with the abscissae of stars on
the Reference Great Circles (RGC), as observed by the satellite.
Then, by calibrating the correlations between the
RGC abscissae, the full covariance matrix $\mathbf{V}$ between
observations allows to find the optimal astrometric parameters. The
method, fully described in van Leeuwen \& Evans (\cite{vleeuwen98}), has been
used with minor differences only.
The calibration of correlation
coefficients has been done on each RGC, the reason being that
significant variations may be found from one orbit to another (Arenou
\cite{arenou97}). This has been done using the theoretical formulae of
Lindegren (\cite{lindegren88}) to which harmonics were added through the
use of cosine transform (Press et al. \cite{press}). Another 
difference from van Leeuwen \& Evans (\cite{vleeuwen98})
comes from the fact that the formal abscissae errors and
correlations have been recalibrated as described in Arenou
(\cite{arenou97}) using the final Hipparcos data, the changes
being at the level of few percent only.

The quantities of interest are the mean parallax $\pi_0$ and the mean
proper motion $\mu_{\alpha 0} \cos \delta_0$, $\mu_{\delta 0}$ of each
cluster centre and the position $\alpha_i, \delta_i$ of each cluster
member $i$. 
When computing the cluster mean parallax, one implicitly assumes that
the  dispersion in individual parallaxes is only due to the measurements
errors. In fact, the depth of the cluster increases the error on the
mean cluster parallaxes by few tenths of mas but should not bias it under
the hypothesis that stars are symmetrically distributed.

In the Hipparcos intermediate data CD-ROM, the abscissae are not given
but their residuals with respect to the main Hipparcos Catalogue
astrometric parameters are given instead. The new residuals on the
abscissae, $\delta{\mathbf{a}}$, with respect to the current iteration value of
($\alpha_i$, $\delta_i$, $\pi_0$, $\mu_{\alpha 0} \cos \delta_0$,
$\mu_{\delta 0}$) are computed. The corrections to these parameters $\delta p_k$
are then found by weighted least-squares, minimizing

\begin{eqnarray*}
(\delta {\mathbf{a}} - \sum_{k=1}^{5}{\partial {\mathbf{a}} \over 
\partial p_k}\delta p_k)^T {\mathbf{V}}^{-1}
(\delta {\mathbf{a}} - \sum_{k=1}^{5}{\partial {\mathbf{a}} \over 
\partial p_k}\delta p_k).
\end{eqnarray*}

Using the partial derivatives ${\partial a_i \over \partial \mu_{\alpha
i} \cos \delta_i}$, ${\partial a_i \over \partial \mu_{\delta i}}$ of
the star $i$ given in the Hipparcos intermediate astrometric data annex,
the partial derivatives of the abscissae with respect to the mean proper
motion (${\mu_{\alpha 0} \cos \delta_0}, {\mu_{\delta 0}})$ are thus
computed using  the linear equations 1 and the relations

{\small
\begin{eqnarray*}
{\partial a_i \over \partial \pi_0} & = &
{\partial a_i \over \partial \pi_i }
{\partial \pi_i \over \partial \pi_0}+
{\partial a_i \over \partial \mu_{\alpha_i} \cos \delta_i }
{\partial \mu_{\alpha_i} \cos \delta_i \over \partial \pi_0}+
{\partial a_i \over \partial \mu_{\delta_i} }{\partial \mu_{\delta_i} 
\over \partial \pi_0}\\
{\partial a_i \over \partial \mu_{\alpha 0} \cos \delta_0} & = &
{\partial a_i \over \partial \mu_{\alpha_i} \cos \delta_i }
{\partial \mu_{\alpha_i} \cos \delta_i \over 
\partial \mu_{\alpha 0} \cos \delta_0}+
{\partial a_i \over \partial \mu_{\delta_i} }
{\partial \mu_{\delta_i} \over \partial \mu_{\alpha 0} \cos \delta_0}\\
{\partial a_i \over \partial \mu_{\delta 0}} & = &
{\partial a_i \over \partial \mu_{\alpha_i} \cos \delta_i }
{\partial \mu_{\alpha i} \cos \delta_i \over 
\partial \mu_{\delta 0}}+
{\partial a_i \over \partial \mu_{\delta_i} }
{\partial \mu_{\delta_i} \over \partial \mu_{\delta_0}}\\
\end{eqnarray*}
}

As part of the least-square procedure, the final covariance matrix
between all the astrometric parameters is also computed.

%
\subsection{Results}
%

%
\renewcommand{\Z}[1]{{\scriptsize#1}}
\begin{table*}[htbp]
\caption{Cluster mean astrometric parameters.}
\begin{center}
\small
\begin{tabular}{lccrrrcc}
\hline
Cluster & NS & uwe & \multicolumn{1}{c}{$\pi$} & $\mu_{\alpha}\cos\delta$ & $\mu_\delta$ 
& $d$ (pc)& $(M-m)_0$\\
   name & NA & NR & \multicolumn{1}{c}{$\sigma_{\pi}$} & $\sigma_{\mu_{\alpha}\cos\delta}$ &
$\sigma_{\mu_\delta}$ &&\\
 &  & & \multicolumn{1}{c}{$\rho_\pi^{\mu_{\alpha}\cos\delta}$} & $\rho_\pi^{\mu_\delta}$ &
$\rho_{\mu_{\alpha}\cos\delta}^{\mu_\delta}$ &&\\
\hline
\bf \object{Coma Ber}     & \bf 30 & 0.97 & \bf 11.49 & \bf -11.38 & \bf -9.05
& $87.0^{+1.6}_{-1.6}$    & $4.70^{+0.04}_{-0.04}$\\[-2pt]
             &   1563  &  15  & 0.21 & 0.23 & 0.12 \\[-3pt]
             &   &    &\Z{-0.13 }&\Z{ 0.06 }&\Z{-0.12}\\
\bf \object{Pleiades}     & \bf 54 & 0.98 &  \bf 8.46 & \bf  19.15 & \bf -45.72
& $118.2^{+3.2}_{-3.0}$   & $5.36^{+0.06}_{-0.06}$\\[-2pt]
               & 2158 &  25  &  0.22 &   0.23 &   0.18\\[-3pt]
             &   &    &\Z{-0.16 }&\Z{-0.07 }&\Z{ 0.21}\\
\bf \object{IC 2391}      & \bf 11 & 0.94 &  \bf 6.85 & \bf -25.06 & \bf 22.73
& $146.0^{+4.8}_{-4.5}$   & $5.82^{+0.07}_{-0.07}$\\[-2pt]
                & 807 &    4 &  0.22 &   0.25 &   0.22\\[-3pt]
             &   &    &\Z{ 0.05 }&\Z{ 0.07 }&\Z{ 0.22}\\
\bf \object{IC 2602}      & \bf 23 & 0.93 &  \bf 6.58 & \bf -17.31 & \bf 11.05
& $152.0^{+3.8}_{-3.6}$   & $5.91^{+0.05}_{-0.05}$\\[-2pt]
             &  1766  &  13  &  0.16 &   0.16 &   0.15\\[-3pt]
             &   &    &\Z{ 0.08 }&\Z{ 0.10 }&\Z{ 0.21}\\
\bf \object{Praesepe}     & \bf 26 & 1.03 & \bf  5.54 & \bf -36.24 & \bf -12.88
& $180.5^{+10.7}_{-9.6}$  & $6.28^{+0.13}_{-0.12}$\\[-2pt]
             &  1126  &   6  &  0.31 &   0.35 &   0.24\\[-3pt]
             &   &    &\Z{-0.22 }&\Z{-0.11 }&\Z{-0.15}\\
\bf \object{NGC 2451}     & \bf 12 & 0.92 &  \bf 5.31 & \bf -22.14 & \bf 15.15
& $188.7^{+7.0}_{-6.5}$   & $6.38^{+0.08}_{-0.08}$\\[-2pt]
              &  908  &   7  &  0.19 &   0.16 &   0.19\\[-3pt]
             &   &    &\Z{ 0.03 }&\Z{ 0.03 }&\Z{-0.03}\\
\bf $\bf \alpha$ Per & \bf 46 & 0.94 &  \bf 5.25 & \bf  22.93 & \bf -25.56
& $190.5^{+7.2}_{-6.7}$   & $6.40^{+0.08}_{-0.08}$\\[-2pt]
             &  2198  &  12  &  0.19 &   0.15 &   0.17\\[-3pt]
             &   &    &\Z{ 0.14 }&\Z{-0.01 }&\Z{ 0.37}\\
\bf \object{Blanco 1}     & \bf 13 & 0.96 &  \bf 3.81 & \bf  19.15 &   \bf 3.21
& $262.5^{+34.3}_{-27.2}$ & $7.10^{+0.27}_{-0.24}$\\[-2pt]
             &   798  &  10  &  0.44 &   0.50 &   0.27\\[-3pt]
             &   &    &\Z{ 0.26 }&\Z{ 0.05 }&\Z{-0.21}\\
\bf \object{NGC 6475}     & \bf 22 & 0.82 &  \bf 3.57 & \bf    2.59 &  \bf -4.98
& $280.1^{+25.7}_{-21.7}$ & $7.24^{+0.19}_{-0.18}$\\[-2pt]
             &  772   &   3  &  0.30 &   0.34 &   0.21\\[-3pt]
             &   &    &\Z{-0.10 }&\Z{ 0.04 }&\Z{-0.12}\\
\bf \object{NGC 7092}     & \bf  8 & 0.92 &  \bf 3.22 & \bf   -7.79 & \bf -19.70
& $310.6^{+30.7}_{-25.7}$ & $7.46^{+0.20}_{-0.19}$\\[-2pt]
             &  589   &   1  &  0.29 &   0.29 &   0.25\\[-3pt]
             &   &    &\Z{-0.07 }&\Z{ 0.03 }&\Z{-0.18}\\
\bf \object{NGC 2232}     & \bf 10 & 0.91 &  \bf 3.08 & \bf   -4.67 &  \bf -3.08
& $324.7^{+41.6}_{-33.1}$ & $7.56^{+0.26}_{-0.23}$\\[-2pt]
             &  497   &   2  &  0.35 &   0.30 &   0.26\\[-3pt]
             &   &    &\Z{-0.07 }&\Z{ 0.03 }&\Z{ 0.05}\\
\bf \object{IC 4756}      & \bf  9 & 0.99 &  \bf 3.03 & \bf   -0.52 &  \bf -5.83
& $330.0^{+59.1}_{-43.5}$ & $7.59^{+0.36}_{-0.31}$\\[-2pt]
             & 522    &   1  &  0.46 &   0.40 &   0.33\\[-3pt]
             &   &    &\Z{ 0.07 }&\Z{ 0.10 }&\Z{ 0.00}\\
\bf \object{NGC 2516}     & \bf 14 & 0.92 &  \bf 2.89 & \bf   -4.04 &  \bf 10.95
& $346.0^{+27.1}_{-23.4}$ & $7.70^{+0.16}_{-0.15}$\\[-2pt]
             & 947    &   4  &  0.21 &   0.22 &   0.20\\[-3pt]
             &   &    &\Z{ 0.10 }&\Z{ 0.05 }&\Z{-0.13}\\
\bf \object{Trumpler 10}  & \bf  9 & 0.97 &  \bf 2.74 & \bf-13.29 &  \bf 7.32
& $365.0^{+43.2}_{-34.9}$ & $7.81^{+0.24}_{-0.22}$\\[-2pt]
             & 702    &   2  &  0.29 &   0.25 &   0.24\\[-3pt]
             &   &    &\Z{ 0.04 }&\Z{ 0.06 }&\Z{ 0.03}\\
\bf \object{NGC 3532}     & \bf  8 & 0.92 &  \bf 2.47 &\bf   -10.84 &  \bf  5.26
& $404.9^{+75.9}_{-55.2}$ & $8.04^{+0.37}_{-0.32}$\\[-2pt]
             & 552    &   5  &  0.39 &   0.38 &   0.37\\[-3pt]
             &   &    &\Z{-0.01 }&\Z{ 0.06 }&\Z{ 0.42}\\
\bf \object{Collinder 140}& \bf 11 & 0.97 &  \bf 2.44 & \bf  -8.52 &  \bf  4.60
& $409.8^{+55.3}_{-43.5}$ & $8.06^{+0.27}_{-0.24}$\\[-2pt]
             & 911 \bf    &   2  &  0.29 &   0.22 &   0.28\\[-3pt]
             &   &    &\Z{ 0.06 }&\Z{ 0.09 }&\Z{ 0.07}\\
\bf \object{NGC 2547}     & \bf 11 & 0.95 &  \bf 2.31 & \bf  -9.28 &   \bf 4.41
& $432.9^{+62.1}_{-48.3}$ & $8.18^{+0.29}_{-0.26}$\\[-2pt]
             & 824    &   3  &  0.29 &   0.31 &   0.24\\[-3pt]
             &   &    &\Z{ 0.10 }&\Z{ 0.15 }&\Z{ 0.06}\\
\bf \object{NGC 2422}     & \bf  9 & 0.97 & \bf  2.01 &  \bf -7.09 &  \bf  1.90
& $497.5^{+135.4}_{-87.7}$& $8.48^{+0.52}_{-0.42}$\\[-2pt]
             &  591   &   1  &  0.43 &   0.35 &   0.28\\[-3pt]
             &   &    &\Z{-0.13 }&\Z{-0.04 }&\Z{ 0.43}\\
\hline
\end{tabular}
\end{center}
$\pi$ and $\sigma_{\pi}$ are in mas, $\mu_{\alpha}\cos\delta$, $\mu_\delta$,
$\sigma_{\mu_{\alpha}\cos\delta}$ and $\sigma_{\mu_\delta}$ are in mas/yr.\\

The notations have the following meaning:\\
NS: number of Hipparcos stars used for the calculation,\\
NA: number of accepted abscissae,\\
NR: number of rejected abscissae,\\
uwe: unit-weight error.\\
\label{table2}
\end{table*}
%

The mean astrometric parameters ($\pi, \mu_\alpha\cos\delta,
\mu_\delta$) and associated standard errors of clusters closer than 500
pc are given in 
Table~\ref{table2}. The unit weights errors are close to 1 but, in general,
slightly smaller. This is possibly because the star to star correlations
between abscissae on the RGCs have not been perfectly calibrated.
However this also suggests that non members have not been erroneously
included and that the cluster depth did not play a significant role. 

The derived distance parameters (distances and distance moduli)
given in Table~\ref{table2}
deserve some further comments. Since the transformation from parallax to
distance or absolute magnitude is not linear, a small bias could be
expected (see Brown et al. \cite{brown}). However, the relative error
$\sigma_\pi/\pi$ is small (between 2 and 20 percent) so the effect is
negligible (between 0.04 percent and 4 percent).

Table~\ref{table3} shows the derived kinematical
parameters ($U$, $V$, $W$) of clusters. They are computed using a solar motion
$(U_{\sun}, V_{\sun}, W_{\sun}) = (10.00, 5.25, 7.17)$ km/s (Dehnen \& Binney
\cite{dehnen98}), with respect to the LSR.

%
\begin{table*}[htbp]
\caption{Cluster mean derived kinematical parameters. The velocity takes 
the solar motion (10.00,5.25,7.17) km/s into account, but not the rotation of 
the LSR}
\begin{center}
\begin{tabular}{l|cc|cccccc|cccccc}
\hline
Cluster & $l$ & $b$ &  $U$ & $\sigma_U$ & $V$ & $\sigma_V$ & $W$ & $\sigma_W$ & 
$\rho_{\pi}^U$ & $\rho_{\pi}^V$ & $\rho_{\pi}^W$ & $\rho_U^V$  & 
$\rho_U^W$ & $\rho_V^W$\\
name & \multicolumn{2}{c|}{degree} & \multicolumn{6}{c|}{km/s} & \multicolumn{6}{c}{percent}\\
\hline
\object{Coma Ber} & 221.28 & 84.03 & 7.82 & 0.09 & -0.31 & 0.12 & 6.62 & 0.25 & 29 & 84 & 4 & 48 & -16 & -10 \\
\object{Pleiades} & 166.62 & -23.57 & 3.65 & 0.45 & -19.12 & 0.69 & -5.85 & 0.35 & 9 & 98 & 75 & -5 & 60 & 64 \\
\object{IC 2391} & 270.36 & -6.89 & -12.92 & 0.75 & -8.33 & 0.25 & 1.18 & 0.24 & 98 & -5 & 64 & -6 & 64 & 2 \\
\object{IC 2602} & 289.63 & -4.89 & 1.56 & 0.37 & -15.01  & 0.30 & 6.88 & 0.12 & 93 & 41 & -10 & 17 & -10 & 13 \\
\object{Praesepe} & 206.07 & 32.34 & -32.43 & 0.88 & -14.99 & 0.44 & -2.02 & 1.51 & 98 & 90 & 99 & 82 & 99 & 88 \\
\object{NGC 2451} & 252.40 & -6.75 & -18.74 & 0.80 & -14.43 & 0.73 & -6.79 & 0.44 & 94 & -40 & 92 & -15 & 90 & -20 \\
\object{$\alpha$ Per} & 146.96 & -7.12 & -5.32 & 0.69 & -20.53 & 0.96 & -0.70 & 0.35 & 80 & 96 & 85 & 61 & 77 & 77 \\
\object{Blanco 1} & 14.95 & -79.30 & -11.68 & 2.51 & -1.78  & 0.90 & -2.35 & 0.54 & 98 & 87 & 91 & 90 & 92 & 85 \\
\object{NGC 6475} & 355.84 & -4.49 & -5.36 & 0.20 & 2.37& 0.44 & 2.04 & 0.71 & 33 & 72 & 80 & 25 & 32 & 42 \\
\object{NGC 7092} & 92.46 & -2.28 & 38.37 & 2.55 & 0.50 & 0.42 & -6.05 & 1.31 & -99 & -14 & 94 & 14 & -93 & -13 \\
\object{NGC 2232} & 214.33 & -7.73 & -5.67 & 0.57 & -6.66 & 0.48 & -4.13 & 1.05 & -32 & 9 & 90 & 31 & -25 & 8 \\
\object{IC 4756} & 36.38 & 5.25 & 35.99 & 0.91 & 13.83& 1.16 & 6.16 & 0.78 & -93 & 92 & 64 & -97 & -58 & 52 \\
\object{NGC 2516} & 273.86 & -15.89 & -7.43 & 1.42 & -18.48& 0.42 & 3.30 & 0.36 & 97 & 31 & -36 & 29 & -32 & -2 \\
\object{Trumpler 10} & 262.82 & 0.63 & -17.25 & 2.65 & -16.62 & 4.97 & -2.48 & 1.16 & 96 & -6 & 93 & 17 & 88 & -10 \\
\object{NGC 3532} & 289.64 & 1.43 & -10.92 & 3.56 & -4.84& 2.66 & 8.32 & 0.83 & 96 & 46 & -16 & 24 & -9 & -11 \\
\object{Collinder 140} & 245.20 & -7.85 & -11.79 & 2.65 & -4.91 & 4.59 & -6.10 & 1.54 & 59 & -20 & 84 & 64 & 83 & 26 \\
\object{NGC 2547} & 264.60 & -8.55 & -8.90 & 2.24 & -5.58 & 1.26 & -6.15 & 1.62 & 97 & -34 & 92 & -30 & 92 & -24 \\
\object{NGC 2422} & 230.98 & 3.13 & -18.26 & 3.14 & -10.50 & 3.25 & -3.65 & 2.71 & 66 & -46 & 94 & 34 & 57 & -49 \\
\hline
\end{tabular}
\end{center}
\label{table3}
\end{table*}
%

Concerning the more distant clusters, the mean cluster parallaxes 
have been computed as described above, under the standard assumption
that members of a given cluster share the common parallax and proper motion.
This concerns 110 clusters more distant than 300 pc with at least 2 Hipparcos 
stars (among which 9 clusters described in Table 2). 
The  parameters of 32 of these clusters containing at least 4 stars observed
by Hipparcos are indicated Table~\ref{table4}. The parameters of the remaining
clusters are not given here though part of them are included in the comparison 
of parallaxes between Hipparcos and groundbased determinations in Sect. \ref{distant}. 

Since the relative parallax error of these distant clusters is 40\% 
on the average, their mean parallaxes are not
useful individually, but rather for statistical studies. Compared to 
these parallaxes, the photometric parallaxes are far more precise.
We could have derived the mean proper motion simultaneously with the 
parallax, but it was prefered to constrain the parallax to its
photometric estimate and to compute the resulting mean proper 
motion (Table~\ref{table4}).  In general, 
these proper motions are close to those obtained without adopting 
the photometric parallax, but this allows to gain one 
degree of freedom.

%
\begin{table}[htbp]
\caption{Mean parameters for all clusters with more than 4 Hipparcos members (\#)
and more distant than 500 pc or with less than 8 Hipparcos members. The proper
motions have been computed constraining the photometric distance estimate
$\pi_{\mathrm P}$. 
This estimate is indicated with its reference: D for Dambis, L for
Loktin \& Matkin, G for Lyng{\aa}, in decreasing order of preference. 
The units are mas for the parallaxes, mas/yr for proper motions; the
correlation coefficient (\%) between $\mu_{\alpha\cos\delta}$ and
$\mu_\delta$ is indicated in the last column.}
\newcommand{\Q}[1]{{\scriptsize$\pm$#1}}
\renewcommand{\Z}[1]{{\tiny#1}}
\begin{center}
\setlength\tabcolsep{3pt}
\begin{tabular}{l|r|r|c|rrr}
\hline
Name&\#&\multicolumn{1}{c|}{$\pi$} &\multicolumn{1}{c|}{$\pi_{\mathrm P}$}& 
\multicolumn{1}{c}{$\mu_{\alpha}\cos\delta$}&
\multicolumn{1}{c}{$\mu_\delta$}&\multicolumn{1}{c}{\Z{$\rho$}}\\
&&\multicolumn{1}{c|}{mas}&\multicolumn{1}{c|}{mas}&
\multicolumn{2}{c}{mas/yr}&\multicolumn{1}{c}{\Z{\%}}\\
\hline
\object{ Cr  121}&13& 1.80\Q{.24}&1.58\Z{D}& -3.88\Q{.16}&  4.35\Q{.19}&\Z{19}\\
\object{ Cr  132}& 8& 1.54\Q{.33}&2.43\Z{G}& -3.57\Q{.24}&  4.16\Q{.31}&\Z{14}\\
\object{ IC 1805}& 4& 1.80\Q{.78}&0.52\Z{D}& -1.14\Q{.71}& -2.29\Q{.62}&\Z{-32}\\
\object{ IC 2944}& 4& 0.56\Q{.43}&0.48\Z{D}& -5.61\Q{.38}&  0.98\Q{.37}&\Z{11}\\
\object{NGC 0457}& 4& 1.55\Q{.58}&0.41\Z{D}& -1.49\Q{.40}& -1.98\Q{.36}&\Z{-39}\\
\object{NGC 0869}& 4& 1.01\Q{.48}&0.54\Z{D}& -0.79\Q{.38}& -1.44\Q{.33}&\Z{-25}\\
\object{NGC 0884}& 5& 0.93\Q{.51}&0.50\Z{D}& -0.77\Q{.42}& -1.87\Q{.35}&\Z{-31}\\
\object{NGC 1647}& 4& 1.09\Q{.80}&2.42\Z{D}& -0.56\Q{.94}& -0.14\Q{.77}&\Z{71}\\
\object{NGC 2244}& 6& 1.37\Q{.56}&0.70\Z{D}& -0.59\Q{.46}&  0.55\Q{.38}&\Z{-12}\\
\object{NGC 2264}& 6& 2.86\Q{.63}&1.39\Z{D}& -0.40\Q{.64}& -4.05\Q{.44}&\Z{27}\\
\object{NGC 2281}& 4& 0.82\Q{.73}&1.89\Z{L}& -2.84\Q{.82}& -7.51\Q{.54}&\Z{17}\\
\object{NGC 2287}& 8& 1.91\Q{.52}&1.53\Z{L}& -4.29\Q{.43}&  0.04\Q{.44}&\Z{ 1}\\
\object{NGC 2467}& 5& 1.79\Q{.65}&0.79\Z{D}& -3.19\Q{.35}&  1.92\Q{.46}&\Z{-5}\\
\object{NGC 2527}& 4& 1.51\Q{.95}&1.65\Z{L}& -6.27\Q{.49}&  8.14\Q{.69}&\Z{11}\\
\object{NGC 2548}& 5& 1.51\Q{.79}&1.51\Z{L}& -0.63\Q{.67}&  0.92\Q{.63}&\Z{-25}\\
\object{NGC 3114}& 6& 1.14\Q{.36}&1.05\Z{D}& -7.77\Q{.39}&  4.15\Q{.31}&\Z{-9}\\
\object{NGC 3228}& 4& 1.39\Q{.50}&1.89\Z{L}&-15.28\Q{.43}&  0.40\Q{.37}&\Z{-9}\\
\object{NGC 3766}& 4& 1.36\Q{.63}&0.59\Z{D}& -7.28\Q{.54}&  1.19\Q{.52}&\Z{28}\\
\object{NGC 4755}& 5& 0.52\Q{.40}&0.53\Z{D}& -4.69\Q{.33}& -1.47\Q{.30}&\Z{36}\\
\object{NGC 5662}& 5& 1.94\Q{.62}&1.39\Z{D}& -5.70\Q{.56}& -7.58\Q{.55}&\Z{-5}\\
\object{NGC 6025}& 4& 0.76\Q{.55}&1.79\Z{D}& -3.63\Q{.47}& -2.87\Q{.53}&\Z{-28}\\
\object{NGC 6087}& 4& 1.30\Q{.61}&1.23\Z{D}& -1.60\Q{.62}& -1.43\Q{.56}&\Z{-10}\\
\object{NGC 6124}& 4& 2.71\Q{.86}&2.15\Z{L}& -1.21\Q{.96}& -1.92\Q{.71}&\Z{-31}\\
\object{NGC 6231}& 6&-0.62\Q{.48}&0.71\Z{D}&  0.04\Q{.47}& -1.94\Q{.34}&\Z{-18}\\
\object{NGC 6405}& 4& 1.69\Q{.52}&2.19\Z{D}& -1.47\Q{.58}& -6.78\Q{.36}&\Z{-28}\\
\object{NGC 6530}& 4& 1.31\Q{.80}&0.79\Z{D}&  1.26\Q{.86}& -2.04\Q{.55}&\Z{-54}\\
\object{NGC 6633}& 4& 2.70\Q{.70}&2.61\Z{L}& -0.09\Q{.60}& -0.39\Q{.51}&\Z{ 7}\\
\object{NGC 6882}& 4& 2.38\Q{.44}&1.68\Z{G}&  2.60\Q{.28}& -9.81\Q{.27}&\Z{-26}\\
\object{NGC 7063}& 4& 2.21\Q{.81}&1.31\Z{L}&  0.43\Q{.52}& -4.24\Q{.56}&\Z{-20}\\
\object{NGC 7243}& 4& 0.43\Q{.61}&1.30\Z{D}&  1.72\Q{.48}& -2.41\Q{.52}&\Z{ 9}\\
\object{Stock 02}& 5& 2.90\Q{.60}&3.30\Z{G}& 15.97\Q{.75}&-13.56\Q{.54}&\Z{-42}\\
\object{  Tr  37}& 6& 1.03\Q{.38}&1.23\Z{D}& -3.75\Q{.35}& -3.48\Q{.33}&\Z{23}\\
\hline
\end{tabular}
\end{center}
\label{table4}
\end{table}
%

%
\section{Systematic errors}\label{systerr}
%
\def\myrho{\rho_{\alpha\cos\delta}^\pi}
%
\subsection{Systematics in the Hipparcos Catalogue}\label{prophip}
%
For the Hipparcos mission, the question of systematic errors has always
been a major issue; it should be remembered that, apart from the higher
number of stars measured, one of the advantages of the Hipparcos data over the
ground-based parallaxes is the uniformity of global astrometry 
observed by a
single instrument. Therefore, during the data reduction special
attention was paid in order to keep the systematics far below
the random errors. A recent study (Makarov, 1998, priv. comm.) shows that
systematic intra-revolution variations of the basic angle or
of the star abscissae, of the order of 4 mas
 through the entire mission, would be needed in
order to produce a 1 mas systematic error of the parallaxes in the 
\object{Pleiades} area. If this had occurred, it would have produced
sizable distortions in other parts of the sky, and consequently a
scatter in parallax measurements much greater than predicted by the
formal errors.

The accuracy and formal precision of the Hipparcos 
data has been 
verified before the delivery of the data (Arenou et al., \cite{arenou95},
\cite{aretal97}, Lindegren \cite{lindegren95}).
Among the available external data of better or 
comparable precision, the comparisons used the best ground-based 
parallaxes, distant stars, distant clusters and Magellanic Cloud 
stars.  In the two latter cases, it should be pointed out that these 
comparisons gave some insight into the property of the parallax errors 
at small angular scale, although the effect of astrometric 
correlations was taken into account only approximately.  In all cases, 
it was shown that, over the whole  catalogue,  not only the zero-point
was smaller than 0.1 mas, 
but also that the formal errors were not underestimated by more than 
$\approx10\%$, this slight underestimation being possibly due to
undetected binaries. In any case, this is far from the $\approx 60\%$
which would be needed for the brighter stars to have 1 mas systematic
errors.

However, the statement of PSSKH that small-scale systematic errors 
may be present in Hipparcos data is not unjustified. Indeed, in a given cluster,
the afore mentioned
correlations between abscissae may be considered as a small error shared by
the stars within a few square degrees.  These errors, probably randomly
distributed over the sky, may thus be regarded as systematics at
small-scale.  However, the method outlined in Sect. \ref{inter}
takes these correlations into account during the computation of the
mean parallax and its associated precision. Then the question is
whether the mean cluster distances and their formal errors appear
statistically biased. The following sections will answer in the negative
using comparisons with previous determinations of cluster parallaxes
and with the help of ad hoc simulations.

%
\subsection{Comparison with previous determinations}
%

The first checking of the mean parallaxes comes from comparison
with previous determinations. The first part of this section mainly
deals with the 7 closest clusters, which have a formal error on
the Hipparcos distance modulus smaller than 0.1 magnitude and can
thus be compared individually with other determinations.
The second part analyses statistically the parallaxes
of the clusters more distant than 300 parsecs.

%
\subsubsection{The closest clusters}
%
Previous cluster distance determinations were mainly derived from
the MSF technique.
With the exception of the \object{Hyades}, where ground-based
trigonometric parallaxes are in excellent agreement with the Hipparcos
ones (see Perryman et al. \cite{perryman}), and the series of papers by
Gatewood et al. (\cite{gatewood90}), Gatewood \& Kiewiet de Jonge (\cite{gatewood94})
and Gatewood (\cite{gatewood95}) (see below), practically no direct determination of
distance exists in the literature.

Distance moduli of the 18 clusters derived from the Hipparcos mean
parallaxes are compared in Table \ref{table5} to those determined by 
\lynga (\cite{lynga87}), Dambis (\cite{dambis}), Loktin \&
Matkin (\cite{loktin}) and Pinsonneault et al. (\cite{pinsonneault}).
Lyng{\aa}'s values, though outdated, are given for
comparison, since Lyng{\aa}'s catalogue of open cluster parameters
has long been the catalogue of reference. These values are
the result of a compilation and do not present any homogeneity. On the
contrary, Loktin \& Matkin (330 clusters) and Dambis (202 clusters)
catalogues are quite homogeneous.
Because the Hipparcos mean distance modulus of the \object{Hyades} is 3.33 $\pm$ 0.1
(Perryman et al. \cite{perryman}),
the distance moduli of Loktin \& Matkin (\cite{loktin}), which are based on a value
of 3.42, are probably systematically overestimated by about 0.1 mag.

Focusing on the 7 nearest clusters for which
the Hipparcos distance modulus errors are smaller than 0.1 magnitude
(and excluding \object{NGC 2451} for the reasons given below),
the following remarks can be done:
\begin{itemize}
\item \object{Coma Ber} and \object{$\alpha$ Per} distance moduli are larger for Hipparcos
than for the other references. Concerning \object{$\alpha$ Per}, 
it should be noticed that the difference between Hipparcos and PSSKH, 
0.17 magnitude, is nearly twice as small as
the difference between PSSKH and Dambis, 0.30 magnitude.
\item The \object{Pleiades} distance modulus is smaller for Hipparcos, but
the difference between Dambis and Hipparcos, 0.11 magnitude, is
in the order of the difference between PSSKH and Dambis, 0.13 magnitude.
\item \object{IC 2391} and 2602 are approximatively at the same distance for Dambis,
Loktin \& Matkin and Hipparcos, but the Hipparcos value is between 
the two others which are discrepant by 0.35 magnitude (0.25 if Loktin
\& Matkin are corrected from the distance modulus of the \object{Hyades}).
\end{itemize}

No systematic differences are, thus, noticeable between Hipparcos distance
moduli and ground-based ones in the sense that there is no general
trend of the Hipparcos distance moduli to be different from all the
MSF distance moduli from all the cited references.
On the contrary, the difference between
Hipparcos and any of these references is of the same order, 0.2 magnitude,
than that between two of these external references.
This behaviour tends to show that the formal errors of distance moduli
derived from the MSF technique are underestimated.
This is not so surprising since MSF
distance moduli depend on the theoretical (or empirical) sequence used,
the metallicity and the redenning chosen and the relations used to
transform ($T_{\rm eff}$, $M_{\rm bol}$) into observable quantities.
For example, an error of 0.1 dex in the metallicity will lead to a
variation of the distance modulus of the order of 0.1 magnitude when using
Johnson $B$,$V$ photometry. And an error of 0.01 magnitude in the reddening
will produce an error of about 0.05 magnitude in $m-M$.

Noticing these discrepancies between the MSF distance moduli, it
would be prudent to consider the Hipparcos data as a good
test of the accuracy of MSF, when the exact chemical composition
of the clusters is not known, and would possibly be a way to give
constraints on this composition.
A review of the consequences of Hipparcos distance moduli on the
MSF technique will be given in the second paper (Robichon et al.
in prep.).

%
\begin{table*}[htbp]
\caption{Hipparcos compared
to previous determinations of cluster distance moduli and redennings.}
\begin{center}
\begin{tabular}{lc|cc|cc|cc|c}
\hline
Cluster & $(m-M)_0$ & $(m-M)_0$ & $E(B-V)$ & $(m-M)_0$ & $E(B-V)$ & $(m-M)_0$ & 
$E(B-V)$ & $(m-M)_0$\\
name & Hipparcos & \multicolumn{2}{c|}{\lynga} & \multicolumn{2}{c|}{Dambis} & 
\multicolumn{2}{c|}{Loktin \& Matkin} & Pinsonneault et al.\\
\hline
\object{Coma Ber}    & $4.70^{+0.04}_{-0.04}$ & 4.49 & 0.00 &               &       &  4.60 & 0.01 & 4.54$\pm$0.04 $^*$\\
\object{Pleiades}    & $5.36^{+0.06}_{-0.06}$ & 5.48 & 0.04 & 5.47$\pm$0.05 & 0.040 &  5.50 & 0.04 & 5.60$\pm$0.04\\
\object{IC 2391}     & $5.82^{+0.07}_{-0.07}$ & 5.92 & 0.01 & 5.74$\pm$0.07 & 0.004 &  6.07 & 0.01 &\\
\object{IC 2602}     & $5.91^{+0.05}_{-0.05}$ & 5.89 & 0.04 & 5.68$\pm$0.05 & 0.038 &  6.07 & 0.05 &\\
\object{Praesepe}    & $6.28^{+0.13}_{-0.12}$ & 5.99 & 0.00 &               &       &  6.26 & 0.02 & 6.16$\pm$0.05\\
\object{NGC 2451}    & $6.38^{+0.08}_{-0.08}$ & 7.49 & 0.04 &               &       &  6.92 & 0.04 &\\
\object{$\alpha$ Per}& $6.40^{+0.08}_{-0.08}$ & 6.07 & 0.09v& 5.94$\pm$0.05 & 0.099 &  6.15 & 0.09 & 6.23$\pm$0.06\\
\object{Blanco 1}    & $7.10^{+0.27}_{-0.24}$ & 6.90 & 0.02 &               &       &       &      &\\
\object{\object{NGC 6475}}    & $7.24^{+0.19}_{-0.18}$ & 6.89 & 0.06 &               &       &       &      &\\
\object{NGC 7092}    & $7.46^{+0.20}_{-0.19}$ & 7.33 & 0.02 &               &       &  7.71 & 0.01 &\\
\object{NGC 2232}    & $7.56^{+0.26}_{-0.23}$ & 7.80 & 0.01 & 7.90$\pm$0.05 & 0.021 &  7.50 & 0.03 &\\
\object{IC 4756}     & $7.59^{+0.36}_{-0.31}$ & 7.94 & 0.20v&               &       &  8.41 & 0.20 &\\
\object{NGC 2516}    & $7.70^{+0.16}_{-0.15}$ & 8.07 & 0.13 & 7.85$\pm$0.05 & 0.111 &  7.86 & 0.10 &\\
\object{Trumpler 10} & $7.81^{+0.24}_{-0.22}$ & 8.09 & 0.06 & 7.80$\pm$0.05 & 0.035 &  7.64 & 0.02 &\\
\object{NGC 3532}    & $8.04^{+0.37}_{-0.32}$ & 8.40 & 0.04 &               &       &  8.23 & 0.04 &\\
\object{Collinder 140}&$8.06^{+0.27}_{-0.24}$ & 7.39 & 0.04 & 7.71$\pm$0.05 & 0.026 &  7.70 & 0.04 &\\
\object{NGC 2547}    & $8.18^{+0.29}_{-0.26}$ & 8.20 & 0.05 & 7.90$\pm$0.10 & 0.054 &  8.16 & 0.04 &\\
\object{NGC 2422}    & $8.48^{+0.52}_{-0.42}$ & 8.37 & 0.08 & 8.13$\pm$0.05 & 0.088 &  8.15 & 0.07 &\\
\hline
\end{tabular}
\end{center}
$^*$ based only on the sequence in the $(M_V, B-V)$ diagram.
v: variable redenning.
\label{table5}
\end{table*}
%

\object{NGC 2451} presents the most discrepant values. The nature of this cluster
was already discussed by R\"oser \& Bastian (\cite{roeser})
and more recently by Platais et al. (\cite{platais96}), Baumgardt \cite{baumgardt}
and Carrier et al. (\cite{carrier}). According to R\"oser \& Bastian
\object{NGC 2451} can be divided into two
different entities. The closest one, at about 220 pc, has a well
defined sequence in the colour-magnitude diagram but presents a large
scatter in proper motion as taken from the PPM catalogue and looks more like a
moving group than like an open cluster. The most distant entity,
situated at about 400 pc, seems to form an open cluster.
Platais et al. (\cite{platais96})
definitively found two clusters \object{NGC 2451}-a and \object{NGC 2451}-b
at 190 and 400 pc utilizing CCD photometry, while Carrier et al.
(\cite{carrier}) confirmed the existence of these two clusters
at 198 and 358 pc from Geneva photometry and the Hipparcos data.
Baumgardt (\cite{baumgardt}) also found \object{NGC 2451}-a at
190 pc from Hipparcos data and supported the existence of \object{NGC 2451}-b
in Hipparcos and ACT data.
Using Hipparcos data alone, \object{NGC 2451}-a ($\pi$=5.30 mas) exhibits a
distinct clump in the vector-point diagram and a well defined peak in
parallax, and has then all the characteristics of an open cluster.
Another peak in the parallax distribution at 2.5 mas, corresponding
probably to \object{NGC 2451}-b, connected with a
concentration in the vector point diagram near ($\mu_\alpha\cos\delta$,
$\mu_\delta$)=(-9, 5) is noticeable. But it is difficult to distinguish
from the field star distribution because both parallax and proper
motion are close to those of field stars.

\object{Pleiades}, \object{Praesepe} and Coma trigonometric parallaxes were obtained from
the ground by Gatewood et al. (\cite{gatewood90}), Gatewood  \& Kiewiet de Jonge
\cite{gatewood94}) and Gatewood (\cite{gatewood95}). In \object{Praesepe},
the mean parallax from
Gatewood (\cite{gatewood94}) is 5.21 $\pm$ 0.8 mas in good agreement
with Hipparcos and MSF values of 
Loktin \& Matkin (\cite{loktin}) and Pinsonneault et al.
(\cite{pinsonneault}). For the \object{Pleiades}, Gatewood et al.
(\cite{gatewood90}) obtained a mean value of 6.6 $\pm$ 0.8 mas, using 5
cluster members. This value is noticeably smaller than both Hipparcos
and MSF values. On the contrary their Coma parallax (Gatewood
et al. \cite{gatewood95}), 13.53 $\pm$ 0.54 mas, is much larger. These
discrepancies may be due to the fact that, although the internal
accuracy of parallaxes are of the order of 1 mas, the zero point, fixed
by 4 field stars for the \object{Pleiades}, 6 for \object{Praesepe} and 8 for Coma,
may be uncertain. There is only one field star in common
with their list, AO 1143 (=HIP 60233). It has a parallax of 2.3 $\pm$ 0.6 in
Gatewood et al. (\cite{gatewood95}) and of 4.27 $\pm$ 0.92 mas in the
Hipparcos Catalogue.

Van Leeuwen \& Evans (\cite{vleeuwen98}) also calculated the mean astrometric
parameters of the \object{Pleiades} and \object{Praesepe} as an example of the use of
Hipparcos intermediate astrometric data. Their method is very similar to
the one presented in this paper as mentioned in Sect. \ref{inter}.
The final obtained values (van Leeuwen \cite{vleeuwen99}), are also 
close to the ones calculated 
in this paper. This is not unexpected since the same abscissae have 
been used in both cases. However different sets of members 
and slight differences in the abscissae formal errors and 
correlations account for the observed differences in the results.

O'Dell et al. (\cite{odell}), used the apparent star
diameters to derive the distances of the \object{Pleiades} and \object{$\alpha$ Per}.
They obtained a distance of 132$\pm$10 pc for the \object{Pleiades} and 
187$\pm$11 pc for \object{$\alpha$ Per}. The value of \object{$\alpha$ Per}
agrees closely with the Hipparcos value while
the distance of the \object{Pleiades} is in agreement with Hipparcos within the
error bars.
The method makes a statistical use of $V\sin i$ of cluster members
associated with their rotational periods and their angular diameters.
Unfortunately, as too few direct angular star diameters are available
for \object{Pleiades} members, a calibration of the diameters
as a function of $V$ and $B-V$ from Hendry et al. (\cite{hendry}) was used.
As for the MSF method, these distances are thus not directly obtained but,
once again, they depend on calibrations which can be biased by several
other parameters like chemical composition or age.

Recently, Chen \& Zhao (\cite{chen}) and Narayanan \& Gould 
(\cite{narayanan}) used purely geometrical methods to derive the distance 
of the \object{Pleiades}. Both methods are based on the hypothesis that members share 
the same space velocity within a small random velocity dispersion of 
a few km/s.

Chen \& Zhao (\cite{chen}) used proper motions and radial velocities
of members to derive the distance and the spatial velocity
of the cluster with a global maximum likelihood procedure.
They obtained a distance of 135.56$\pm$0.72 pc. The tiny error
bar seems dubious. In addition, they used the proper motions of Hertzsprung
(\cite{hertzsprung}) which are only relative. The zero point of
the proper motions is not given. From their resulting space velocity,
the components of the proper motions
($\mu_\alpha\cos\delta$,$\mu_\delta$) can be estimated to be (21.50, -33.04).
The component in declination is quite different
from the Hipparcos mean proper motion of the cluster. Moreover, the differences
between Hertzsprung's proper motions and the ACT catalogue proper motions
(Urban et al. \cite{act}) show very significant dependencies with magnitudes
and coordinates. This suggests biases in the Herstzsprung catalogue
of the order of few mas/yr. No discussion on proper-motion biases, neither
on the discrepant value of the mean proper motion, is given by
Chen \& Zhao (\cite{chen}).

Narayanan \& Gould (\cite{narayanan}) used the gradient of radial velocities
to derive a \object{Pleiades} mean distance of 130.7$\pm$11.1 pc, in agreement
with Hipparcos within the error bars.
Their set of 154 individual radial velocities is a compilation of CORAVEL
measurements taken from the same references as those of Sect. \ref{selection}. 
They used a mean proper motion of ($\mu_\alpha\cos\delta, \mu_\delta)
=(19.79, -45.39)$ computed as an average of 65 Hipparcos members.
They explained the difference with the Hipparcos mean parallax
by small scale correlations between individual Hipparcos parallaxes, 
greater than those described above. However, following their 
arguments, if the mean Hipparcos parallax is biased, then the mean proper 
motion could also be biased. The fact that 
Narayanan \& Gould use an average of the Hipparcos proper motions 
could be a problem since a variation of 1 mas/yr in $\mu_\delta$, for instance, 
modifies the mean distance by about 2.5 pc.

In order to analyse the radial-velocity gradient method, 
a new selection of radial velocity members was done. All the members 
with a CORAVEL radial velocity were considered. The spectroscopic
binaries were rejected when they had no orbital
solution as well as all stars with less than 3 measurements (and which thus could
also be non detected spectral binaries). 133 stars were selected on this basis. 
Their mean distance
is 133.8$\pm$9.3 pc using the radial-velocity gradient method and
the mean values of the centre, mean radial velocity and proper motion
indicated in Tables \ref{table1} and \ref{table2} respectively. This distance
confirms the result of Narayanan \& Gould (\cite{narayanan}).
However some doubts can be casted upon the assumption that all the members
share the same space velocity and are at the same distance.
Adopting the same notations as Narayanan \& Gould (\cite{narayanan}), let
$V_{r,i}$ be the observed radial velocity of a member $i$, $\bf n_i$
the unit vector pointing in its line of sight and $V_r$, \mbox{\boldmath$\mu$} and $\bf n$
the mean radial velocity, mean proper motion and direction of the cluster center.
Figure 2 of Narayanan \&
Gould (\cite{narayanan}) shows the difference between $V_{r,i} - V_r(\bf n.n_i)$
versus \mbox{\boldmath$\mu$\unboldmath{$\bf .n_i$}}. The slope of the linear
regression of these points
gives directly the distance of the cluster. The most weighty points are then those
with the most extreme values of the proper-motion projection on the line of sight, i.e.
the most distant members from the cluster centre parallel to the proper motion
direction. The cluster distance derived selecting only the 27 stars satisfying
\mbox{$|$\boldmath$\mu$\unboldmath{$\bf .n_i$}$| > 7$} is 145$\pm$11 pc while it is
100$\pm$16 pc when using the 106 other members. This behaviour is quite puzzling.
If the CORAVEL data are free from any bias, this could indicate that the spatial structure
of the cluster is not symmetrical or that the member velocity dispersion is not
uniform, due to tidal distortion by the galactic potential for example.
Nevertheless, investigations need to be carried out and would probably be the
subject of a further paper.

Summarizing this paragraph leads to two distance estimates for the \object{Pleiades}.
The Hipparcos one around 120 pc (this paper and van Leeuwen \cite{vleeuwen99})
and a group of other values around 130 pc (PSSKH, O'Dell et al. \cite{odell},
Chen \& Zhao \cite{chen}, and Narayanan \& Gould \cite{narayanan}), 
part of them being compatible with the Hipparcos result within the error bars.

%
\subsubsection{Statistical properties of distant cluster mean 
parallaxes}\label{distant}
%
The MSF method may be used efficiently for distant (e.g. $>300$ pc) 
clusters, since in this case, for a given absolute magnitude error,
the photometric parallax error becomes far smaller than the Hipparcos 
parallax error.  Even a systematic absolute magnitude shift would 
only produce a slight asymmetry on the distribution of differences between
Hipparcos and MSF parallaxes; this may be seen when Hipparcos is compared to
Loktin \& Matkin (\cite{loktin}) distance moduli, in Arenou \& Luri
(\cite{arenou99}).

Since these distant clusters are much more concentrated on the sky 
than the nearby clusters, the effect of angular correlations should also
be more obvious.  If systematic errors were present in the Hipparcos 
mean cluster parallaxes, then they would show up as either a 
systematic offset when cluster parallaxes are compared to photometric 
parallaxes, or as a scatter not accounted for in the formal errors.  
On the contrary, the errors on the normalized parallax differences 
appear normally distributed, the Gaussian (0,1) null hypothesis being 
compatible with the observations.

A further piece of evidence that the RGC correlations (and consequently the 
formal error on the mean cluster parallaxes) seem to have been 
correctly taken into account is shown Table~\ref{nombreamas}, where 
the mean parallaxes are compared with those deduced from Dambis 
(\cite{dambis}).  This reference was chosen because the formal error 
of the photometric parallaxes is indicated. Therefore, the statistical 
properties of mean cluster parallaxes may be safely studied.

For the 66 clusters more distant than 300 pc, with at least two 
members and a Dambis distance modulus, the normalized 
differences between Hipparcos and Dambis parallaxes have been 
calculated. Then, the mean formal error $\langle\sigma_\pi\rangle$
and the unit-weight error (RMS error of the normalized differences) in several
groups of clusters containing the same number of Hipparcos stars, have been
computed (Table \ref{nombreamas}).
If systematic errors were present, the RMS error should increase
with the number of stars in each cluster (since the mean 
formal error $\langle\sigma_\pi\rangle$ decreases).
No such trend has been found and the random
errors are mainly responsible of the departure from the expected 
value (equal to 1 if the formal parallax errors are realistic). The average
unit-weight error, 1.15, is not that bad since
the membership in these distant clusters is not firmly determined.
There is then no room for 1 mas systematic error or in only very few clusters.

%
\begin{table}[htbp]
\caption{RMS normalized differences between cluster parallaxes and 
Dambis photometric parallaxes as a function of number of Hipparcos 
stars in each cluster.}
\begin{center}
\begin{tabular}{r|r|r|r}
\hline
\# of & \# of &$\langle\sigma_\pi\rangle$& RMS\\
members & \multicolumn{1}{c|}{clusters} & (mas) &\\
\hline
      2& 28 & 1.03 & 1.00\\
      3& 11 & 0.80 & 1.25\\
      4& 12 & 0.60 & 1.26\\
      5&  4 & 0.55 & 0.98\\
      6&  6 & 0.48 & 1.57\\
$\geq 9$& 5 & 0.29 & 0.99\\
\hline
\end{tabular}
\end{center}
\label{nombreamas}
\end{table}
%
The estimation of the formal error of the mean parallax based on distant
clusters seems statistically realistic. There is then no reason to suspect
the presence of a problem on closer clusters, because the error on the parallax is 
independent from the parallax itself (Arenou et al. \cite{arenou95}).  Concerning 
the \object{Pleiades}, this suggests that the formal parallax error has 
been correctly estimated.  The \object{Pleiades} could of course be at $4\sigma$ 
from the true parallax, but this is improbable, except if this is one 
special case where the small-scale correlations have been 
severely underestimated.

No reason however has been found, which could justify this hypothesis.
For instance, one way of testing the way the
small-scale correlations were taken into account is to study the 
variations of astrometric parameters with the angular distance between stars.
\object{Pleiades} stars have been grouped in six bins of nine stars with increasing
distances from the centre and the mean astrometric parameters for each
bin are given in Table \ref{distpleiad}. 
All values are compatible with the adopted mean values, and no 
significant trend appears for $\pi$ or $\mu_\alpha \cos\delta$.
Concerning $\mu_\delta$, the
last bin (containing the 9 farthest stars from the cluster centre) 
is at 3.2 $\sigma$ from the cluster mean value. If these 9 stars were
rejected, the new mean values of the astrometric parameters remain
compatible with the adopted values, but the last bin would then be at more 
than $2\sigma$ in $\mu_{\alpha\cos\delta}$ and $4\sigma$ in $\mu_\delta$.

%
\begin{table}[htbp]
\caption{Mean astrometric parameters on subsamples of the \object{Pleiades} 
(6 bins of 9 stars) selected by increasing distances from the cluster 
centre. The average angular distance $<d>$ from the cluster centre is indicated}
\begin{center}
\begin{tabular}{c|c|c|c|c}
\hline
bin &$<d>$ & $\pi$ & $\mu_{\alpha\cos\delta}$ & $\mu_\delta$ \\
\# &\degr & mas & mas/yr & mas/yr\\
\hline
1 & 0.35 & $8.30\pm 0.46$ &  $19.73\pm 0.43$ & $-44.87\pm 0.32$ \\
2 & 0.67 & $9.07\pm 0.42$ &  $18.59\pm 0.42$ & $-45.05\pm 0.32$ \\   
3 & 1.31 & $8.89\pm 0.45$ &  $19.35\pm 0.46$ & $-45.59\pm 0.36$ \\   
4 & 1.96 & $7.37\pm 0.58$ &  $18.97\pm 0.59$ & $-45.58\pm 0.45$ \\
5 & 2.84 & $8.65\pm 0.49$ &  $18.53\pm 0.51$ & $-45.45\pm 0.41$ \\
6 & 4.24 & $8.17\pm 0.43$ &  $19.98\pm 0.51$ & $-46.94\pm 0.38$ \\ \hline
all & 1.90 & $8.46\pm 0.22$ & $19.15\pm 0.23$ & $-45.72\pm 0.18$ \\ \hline
\end{tabular}
\end{center}
\label{distpleiad}
\end{table}
%
 
No definitive explanation has been found to explain this behaviour. 
However, if we add to this problem what has yet been
noticed about the radial velocities, and if the effect on parallaxes 
shown by Narayanan \& Gould is not interpreted as systematics, 
there are indications that the
spatial and/or kinematical distributions of the \object{Pleiades} are not as regular 
as expected. This is possibly an explanation to the so-called \object{Pleiades} anomaly.

%
\subsection{The effect of $\myrho$}
%
According to PSSKH, systematic errors, on the order of 1 mas
and thus far greater than the mean random error, are present
in the Hipparcos Catalogue. They would be due to the existing
correlations between right ascension and parallax for stars within a
small angular region. PSSKH have shown that there is a trend in $\pi$ vs
$\myrho$ for the \object{Pleiades}, where the most luminous stars near the cluster
centre, and with the highest $\myrho$, are those which raise the average
parallax above that expected from MSF. In view of their results,
PSSKH cautioned the users of Hipparcos data
for the stars with high $\myrho$. 

This section shows that, on the contrary, no bias on the parallax can
be attributed to $\myrho$ neither on large scale nor on small scale.

%
\subsubsection{Behaviour over the whole sky}
%
Using the whole Hipparcos Catalogue, stars more distant than 500 pc according
to their $uvby\beta$ photometry have been selected. Taking into account this selection bias
as described in Arenou et al. (1995), Section 4, the
zero-point of Hipparcos parallaxes is found to be $-0.09\pm 0.14$ mas
for 74 stars with $\myrho> 0.3$, with an unit-weight error of parallax
$0.94\pm0.10$, whereas the same computation with no restriction on
$\myrho$ gives $-0.05\pm 0.05$ mas. Thus high $\myrho$ do not 
seem to play a special role on the parallax of individual stars.

However, this does not exclude possible effects
at small angular scales. For this purpose, the mean parallaxes for 
distant clusters have been compared to their photometric counterpart.
Using the 66 clusters more distant than 300 pc, 
the average difference between cluster parallaxes, 
(Hipparcos $-$ Dam\-bis), is indicated Table~\ref{rhoamas} in 7 
quantiles of 9-10 clusters according to the average $\myrho$.
Although some significant departures from 0 are present when individual
stars are used (Arenou \& Luri \cite{arenou99}), the correlation 
$\myrho$ does not seem to influence the mean cluster parallaxes. There is 
only one significant bin, at $\myrho\approx -0.2$, which is mainly due 
to one cluster, \object{NGC 6231}, where all stars have a negative parallax.
The \object{Pleiades} being in the last bin, a 1 mas error due to $\myrho$
would be improbable. 

%
\begin{table}[htbp]
\caption{Errors on mean Hipparcos cluster parallax (mas)
as a function of the cluster average $\myrho$.}
\begin{center}
\begin{tabular}{r|r}
\hline
$\langle\myrho\rangle$&$\langle\pi_{\mathrm{Hip}}-\pi_{\mathrm{Dambis}}\rangle$\\
\hline
-0.34 & $-0.34\pm 0.32$\\
-0.20 & $-0.53\pm 0.21$\\
-0.10 & $ 0.10\pm 0.32$\\
-0.04 & $-0.29\pm 0.30$\\
 0.03 & $ 0.17\pm 0.18$\\
 0.11 & $-0.04\pm 0.23$\\
 0.32 & $ 0.29\pm 0.31$\\
\hline
\end{tabular}
\end{center}
\label{rhoamas}
\end{table}
%

%
\subsubsection{Small scale effect: the \object{Pleiades}}
%
PSSKH found a slope of 3.04 $\pm$ 1.36 mas when computing a linear 
regression between $\myrho$ and $\pi$ of their \object{Pleiades} members.  For 
the members determined in this study, a slope of 1.95 $\pm$ 0.99 mas 
has been obtained.  
PSSKH interpreted this slope as the signature of an Hipparcos 
systematic error.  It should however be remembered that a 
correlation is not always a causality.  In the present case, the 
slope comes partially from the fact that the members in the central 
part of the cluster share the same RGCs.  This implies
that the individual values of the parallax are 
correlated.  This also implies that the $\myrho$ values are similar 
since the distribution of time on the parallactic ellipses are nearly 
the same.  Due to the scanning law of the satellite in this area, the 
correlations are all around 0.3.  But there are no reason for believing 
that an unbiased value of the mean parallax can be derived using 
$\myrho=0$.  Two kinds of Monte-Carlo simulations 
have been done in order to assess these points.

First, using the assumed mean Hipparcos parallax and proper 
motion (given in Table \ref{table2}) of the \object{Pleiades}, simulated 
abscissae have been generated, using the complete covariance matrix of 
these observations for the cluster.  For each star, an 
astrometric solution has been performed.  For each simulated \object{Pleiades}, 
the $\myrho$ and $\pi$ of each star member and a mean value of $\pi$ 
derived from the intermediate data are computed.  The mean slope 
between $\myrho$ and $\pi$ over the simulations spread from -3.9 mas 
to 2.9 mas with a mean value $-0.12\pm 0.14$ mas.  The mean value of the 
mean parallaxes is $8.45\pm 0.25$ mas.  Keeping only the simulated 
\object{Pleiades} with a slope greater than 2 (less than 10\% of the 
simulations), the mean parallax is $8.55\pm 0.13$ mas.  This fully 
demonstrates that the weight of the stars with a large $\myrho$ do 
not bias the mean parallax value.

Secondly, the $\myrho$ correlation appears for a star if the 
repartition of Hipparcos Reference Great Circles for this star is 
asymmetrical with regard to the position of the Sun (see chapter 3.2 
of the Hipparcos Catalogue \cite{hip}).
In the case of \object{Pleiades} stars, the RGCs are splitted 
into two groups of 2.5 months over the year, due to the scanning law 
of the satellite.  The first group is centred on mid February and 
contains twice as many RGCs as the second group which is centred on mid 
August.  To reduce the $\myrho$ values, new reductions computing both 
individual and mean cluster astrometric parameters were carried out,
while rejecting randomly half of the RGCs of the first group.  As expected, 
the $\myrho$ values became equal to zero on the average ($-0.01\pm 
0.02$), but the mean parallax still remains quite the same on the 
average ($8.40\pm 0.12$ mas), the slope between individual $\myrho$ and 
$\pi$ remaining positive.

One can conclude, from these two groups of simulations, that the mean 
values of the cluster parallax do not depend on the correlations 
$\myrho$.

%
\subsection{Other possible effects}
%
%
\subsubsection{Bad RGCs}
%
For a normal RGC the individual precision on a star abscissa residual
is 3 mas on average, the mean value being 0.
If for some reason bad RGCs had a large weight, the mean parallax 
could be biased.  For example, in the \object{Pleiades}, the mean parallaxes 
derived when removing all the abscissae of the RGC 221 or 1519 are, 
respectively, 8.24$\pm$0.23 and 8.56$\pm$0.23.  These are the extreme 
cases, for which a convergence of factors are responsible: the large
number of stars observed on these RGCs, the high value of the partial 
derivative ${\partial a \over \partial \pi}$, the high parallactic 
factors at the time of observation
and the good accuracy of the abscissae.  The 
influence of the other RGCs is, in most cases, smaller than 0.05 
mas.  Anyway, except perhaps RGC 674, which has a lot of outliers,
there is no indication that any particular RGC should be 
removed. And the mean astrometric parameters remain the same when
discarding RGC 674.

%
\subsubsection{Binarity}
%
The possibility that systematic errors could originate from undetected
binarity has also been checked for the \object{Pleiades} case. Apart from binary
stars flagged as such by Hipparcos, and rejected in the solutions given
in the previous section, a solution has also been performed where all the
ground-based (spectroscopic or visual) binaries (20 stars) were
rejected. The resulting average parallax (8.50$\pm$0.26) is not
significantly different from the adopted solution. 

In fact, excluding the rare cases where the period of the binary is 
about one year, no parallax bias due to unknown binarity is expected.  
To assess this point, a simple test has been done: using the stars 
given in the orbital solutions of the Hipparcos DMSA/O annex, and 
computing a single star solution instead does not change significantly 
their parallax estimate.  Since the binarity of these stars was known, 
undetected binaries (which implies a much smaller astrometric 
perturbation) are thus less likely to produce a significant effect on 
the parallax.

%
\section{Conclusions}
%
Open clusters have been used for a long time to calibrate the main 
sequence in the Hertzsprung-Russell diagram as a function of age and 
chemical composition and define one of the first steps in the distance 
scaling of the Universe via photometric parallaxes.

The Hipparcos Catalogue allows, for the first time, to 
determine, without any physical assumption, the distances of the 
nearest open clusters presented here, and the locations of the 
cluster sequences in the HR diagram, which will be studied in detail 
in a further paper.

A new selection of members, based on Hipparcos main Catalogue data, in 
the 9 clusters closer than 300 pc (except the \object{Hyades}) and in 9 rich 
clusters between 300 and 500 pc, has been carried out. To 
these nearby clusters, a selection of 32 more distant clusters  
with at least 4 Hipparcos stars has also been added.

New mean astrometric parameters have been compu\-ted using Hipparcos 
intermediate data, taking account of the star to star correlations. 
The precisions are better than 0.5 mas for parallaxes and 0.5 mas/yr for
proper motions.
For the most distant clusters the relative precision of the mean parallax
is not as good but they may be used for statistical purposes. Proper
motions, computed using the photometric parallaxes, may also be useful e.g. 
for the kinematical study of young stars.

Extensive tests have been applied, on distant clusters as well as on 
the \object{Pleiades}, which show that no obvious systematic errors seem to be 
present in the obtained results, and that the computed precisions are 
representative of the true external errors.  This should allow in turn 
to improve the MSF distance moduli and to obtain reliable 
estimates of their external errors.

%
\begin{acknowledgements}
%
We thank Dr F. Mignard for his comments about the \object{Pleiades} Hipparcos data.
We are grateful to Dr B. Miller for having improved the style of the paper
and to Dr T. de Zeeuw and J. de Bruijne for helpful comments.
Extensive use has been made of the Simbad database, operated at CDS, 
Strasbourg, France, and of the Base Des Amas (WEBDA).
%
\end{acknowledgements}
%

%

%
\section*{Appendix: Hipparcos cluster members}\label{app}
%
The appendix lists the Hipparcos stars selected as members in the nearby
clusters. Table \ref{table9} contains the members seen as multiple by
Hipparcos and not used in the mean parameter calculation (H59 = C, O, G,
V or X), while table \ref{table10} gives the numbers of single Hipparcos
stars used.

%
\begin{table*}[htbp]
\caption{Hipparcos cluster members flagged as multiple stars in HIP and 
not used for the calculation of the cluster mean astrometric parameters}
\begin{center}
\footnotesize
\begin{tabular}{lclll|lclll}
Cluster name & H59 & \multicolumn{3}{c}{Hipparcos number} & 
Cluster name & H59 & \multicolumn{3}{c}{Hipparcos number}\\
\hline
\object{Coma Ber}     & C & 60525 &       &       & \object{NGC 6475}     & C & 87218 & 87567 \\
\object{Pleiades}     & C & 17572 & 17923 &       &              & V & 87063 \\
	     & O & 17694 & 17847 &       & \object{NGC 7092}     & C & 106262 \\
             & G & 18559 &       &       &              & G & 106170 & 106409 \\
\object{IC 2391}      & C & 42216 & 42715 &       & \object{NGC 2232}     & X & 30076 \\
\object{IC 2602}      & C & 52116 & 52171 & 52815 &              & C & 30535 \\
             &   & 53330 &       &       & \object{NGC 2516}     & C & 38416 & 38416 & 38966\\
             & O & 52419 &       &       &              &   & 39195 & 39562 \\
             & X & 51794 & 43044 &       & \object{Trumpler 10}  & C & 43085 & 43087 & 43680\\
\object{Praesepe}     & C & 42497 &       &       &              &   & 43688\\
             & G & 42542 &       &       & \object{NGC 3532}     & C & 54184 & 54809 \\
\object{NGC 2451}     & C & 37322 &       &       & \object{NGC 2547}     & C & 39479 \\
\object{$\alpha$ Per} & C & 16244 &       &       &              & \\
\hline
\end{tabular}
\end{center}
\bigskip
\label{table9}
\caption{Hipparcos cluster members used for the calculation of the
cluster mean astrometric parameters}
\begin{center}
\footnotesize
\begin{tabular}{lrrrrrrrrrr}
Cluster name & \multicolumn{10}{c}{Hipparcos number}\\
\hline
\object{Coma Ber}    &   59364 &  59399 &  59527 &  59833 &  59957 &  60025 &  60063 &  60066 &  60087 & 60123\\
            &   60206 &  60266 &  60293 &  60304 &  60347 &  60351 &  60406 &  60458 &  60490 & 60582\\
            &   60611 &  60649 &  60697 &  60746 &  60797 &  61071 &  61074 &  61147 &  61295 & 61402\\
\object{Pleiades}    &   16217 &  16407 &  16423 &  16635 &  16639 &  16753 &  16979 &  17000 &  17020 & 17034\\
            &   17043 &  17044 &  17091 &  17125 &  17225 &  17245 &  17289 &  17316 &  17317 & 17325\\
            &   17401 &  17481 &  17489 &  17497 &  17499 &  17511 &  17527 &  17531 &  17547 & 17552\\
            &   17573 &  17579 &  17583 &  17588 &  17608 &  17625 &  17664 &  17692 &  17702 & 17704\\
            &   17729 &  17776 &  17791 &  17851 &  17862 &  17892 &  17900 &  17999 &  18050 & 18091\\
            &   18154 &  18431 &  18544 &  18955 &\\
\object{IC 2391}     &   42274 &  42374 &  42400 &  42450 &  42459 &  42504 &  42535 &  42702 &  42714 & 42726\\
            &   43195 &\\
\object{IC 2602}     &   50102 &  50612 &  51131 &  51203 &  51300 &  51576 &  52059 &  52132 &  52160 & 52221\\
            &   52261 &  52293 &  52328 &  52370 &  52502 &  52678 &  52701 &  52736 &  52867 & 53016\\
            &   53913 &  53992 &  54168 &\\
\object{Praesepe}    &   41788 &  42106 &  42133 &  42164 &  42201 &  42247 &  42319 &  42327 &  42436 & 42485\\
            &   42516 &  42518 &  42523 &  42549 &  42556 &  42578 &  42673 &  42705 &  42766 & 42952\\
            &   42966 &  42974 &  42993 &  43050 &  43086 &  43199 &\\
\object{NGC 2451}    &   36653 &  37297 &  37450 &  37557 &  37623 &  37666 &  37697 &  37752 &  37829 & 37838\\
            &   37982 &  38268 &\\
\object{$\alpha$ Per} &  14697 &  14853 &  14949 &  14980 &  15040 &  15160 &  15259 &  15363 &  15388 & 15404\\
            &   15420 &  15444 &  15499 &  15505 &  15531 &  15556 &  15654 &  15770 &  15819 & 15863\\
            &   15878 &  15898 &  15911 &  15988 &  16011 &  16036 &  16047 &  16079 &  16118 & 16137\\
            &   16147 &  16210 &  16318 &  16340 &  16403 &  16426 &  16430 &  16455 &  16470 & 16574\\
            &   16625 &  16782 &  16826 &  16880 &  16966 &  16995 &\\
\object{Blanco 1}    &     163 &    212 &    232 &    257 &    328 &    349 &    389 &    395 &    477 &   512\\
            &     585 &    653 &     77 &\\
\object{NGC 6475}    &   87102 &  87134 &  87230 &  87240 &  87360 &  87460 &  87472 &  87516 &  87529 & 87560\\
            &   87580 &  87616 &  87624 &  87656 &  87671 &  87686 &  87698 &  87722 &  87785 & 87798\\
            &   87844 &  88247 &\\
\object{NGC 7092}    &  105658 & 105659 & 105955 & 106270 & 106293 & 106297 & 106329 & 106848 &\\
\object{NGC 2232}    &   30197 &  30356 &  30595 &  30660 &  30700 &  30758 &  30761 &  30772 &  30789 & 31101\\
\object{IC 4756}     &   90958 &  90990 &  91171 &  91299 &  91312 &  91437 &  91513 &  91870 &  91909 &\\
\object{NGC 2516}    &   38226 &  38310 &  38433 &  38536 &  38739 &  38759 &  38783 &  38906 &  38994 & 39070\\
            &   39073 &  39386 &  39438 &  39879 &\\
\object{Trumpler 10} &   42477 &  42939 &  43055 &  43182 &  43209 &  43240 &  43285 &  43326 &  43450 &\\
\object{NGC 3532}    &   54147 &  54177 &  54197 &  54237 &  54266 &  54294 &  54306 &  54337 &\\
\object{Collinder 140}&  35432 &  35641 &  35700 &  35761 &  35795 &  35822 &  35855 &  35905 &  36038 &  36045\\
             &  36217\\ 
\object{NGC 2547}    &   39679 &  39759 &  39988 &  40011 &  40016 &  40024 &  40059 &  40336 &  40353 & 40385\\
            &   40427 &\\
\object{NGC 2422}    &   36717 &  36773 &  36967 &  36981 &  37015 &  37018 &  37037 &  37047 &  37119 &\\
\hline
\end{tabular}
\end{center}
\label{table10}
\end{table*}
%

\begin{thebibliography}{}
%
\bibitem[1997]{arenou97}
Arenou F., 1997, ESA SP-1200, vol.3, chap. 17

\bibitem[1999]{arenou99}
Arenou F., Luri X., 1999, In `Harmonizing Cosmic Distance Scales in a 
Post-Hipparcos Era', D. Egret \& A. Heck eds, ASP Conf. Series

\bibitem[1995]{arenou95}  
Arenou F., Lindegren L., Fr{\oe}schl{\'e} M.,
et al., 1995, A\&A 304, 52

\bibitem[1997]{aretal97}
Arenou F., Mignard F., Palasi J., 1997, ESA SP-1200, vol.3, chap. 20

\bibitem[1998]{baumgardt}
Baumgardt H., 1998, A\&A 340, 402

\bibitem[1997]{brown}
Brown A.G.A., Arenou F., van Leeuwen F., Lindegren L., Luri X., 1997, 
Hipparcos Venice'97, ESA SP-402, p. 63

\bibitem[1998]{carrier}
Carrier F., Burki G., Richard C., 1998, A\&A 341, 469

\bibitem[1997]{chen}
Chen L., Zhao J.L., 1997, Chin. Astron. \& Astrophys., 21, 433 

\bibitem[1999]{dambis}
Dambis A. K., 1999, Astronomy Letters, in press

\bibitem[1998]{dehnen98}
Dehnen W., Binney J.J., 1998, MNRAS 298, 387

\bibitem[1995]{web}
Duflot M., Figon P., Meyssonnier N., 1995, A\&AS 114, 269 

\bibitem[1997]{hip}
ESA, The Hipparcos Catalogue, 1997, Volume 1, Introduction and Guide 
to the Data, ESA SP-1200

\bibitem[1995]{gatewood95}
Gatewood G., 1995, ApJ 445, 712 

\bibitem[1994]{gatewood94}
Gatewood G., Kiewiet de Jonge J., 1994, ApJ 428, 166 

\bibitem[1990]{gatewood90}
Gatewood G., Castelaz M., Han I., Persinger T., Stein J., 1990, 
ApJ 364, 114 

\bibitem[1993]{hendry}
Hendry M.A., O'Dell M.A., Collier Cameron A., 1993, MNRAS 265, 983

\bibitem[1947]{hertzsprung}
Hertzsprung E., 1947, Ann. Sterrew. Leiden 19, 3

\bibitem[1997]{jeffries}
Jeffries R.D., Thurston M.R., Pye J.P., 1997, MNRAS 287, 350

\bibitem[1999]{vleeuwen99}
van Leeuwen F., 1999, In `Harmonizing Cosmic Distance Scales in a 
Post-Hipparcos Era', D. Egret \& A. Heck (eds.), ASP Conf. Series

\bibitem[1998]{vleeuwen98}
van Leeuwen F., Evans D., 1998, A\&AS 130, 157

\bibitem[1988]{lindegren88}
Lindegren L., 1988, In `Scientific aspects of the Input Catalogue 
Preparation II', January 1988, Sitges, J. Torra, C. Turon, (eds.)

\bibitem[1995]{lindegren95}
Lindegren L., 1995, A\&A 304, 61

\bibitem[1994]{loktin}
Loktin A.V., Matkin N.V., 1994, Astron. Astrophys. Trans. 4, 153

\bibitem[1987]{lynga87}
Lyng{\aa} G., 1987, Catalogue of open cluster data, 5th edition, 
Centre de Donn\'ees Stellaires, Strasbourg

\bibitem[1995]{mermilliod95}
Mermilliod J.C., 1995, In `Information and On-Line Data in Astronomy',
D. Egret \& M.A. Albrecht (eds.), Kluwer, 127

\bibitem[1998]{mermilliod98}
Mermilliod J.C., 1998, In `Very low-mass stars and brown dwarfs in
stellar clusters and associations'
R. Rebolo \& M.R. Zapatorio Osorio (eds.), Cambridge Univ. Press

\bibitem[1989]{mermilliod89}
Mermilliod J.C., Turon C., 1989, In `The Hipparcos Mission. 
Prelaunch Status.', ESA SP-1111, Vol. 2, 177 

\bibitem[1990]{mermilliod90}
Mermilliod J.C., Weis E.W., Duquennoy A., Mayor M., 1990,
 A\&A 235, 114

\bibitem[1997a]{mermilliod97a}
Mermilliod J.C., Turon C., Robichon N., Arenou F., 
Lebreton Y., 1997a, ESA SP-402, 643

\bibitem[1997b]{mermilliod97b}
Mermilliod J.C., Bratschi P., Mayor M., 1997b, A\&A 320, 74

\bibitem[1999]{narayanan}
Narayanan V.K., Gould A., 1999, AJ submitted

\bibitem[1994]{odell}
O'Dell M.A., Hendry M.A., Collier Cameron A., 1994, MNRAS 268, 181

\bibitem[1998]{perryman}
Perryman M.A.C. , Brown A.G.A., Lebreton Y.,
et al., 1998, A\&A 331, 81 

\bibitem[1998]{pinsonneault}
Pinsonneault M.H., Stauffer J., Soderblom D.R., King J.R., 
Hanson R.B., 1998, ApJ 504, 170

\bibitem[1996]{platais96}
Platais I., Kozhurina-Platais V., Barnes S., Horch E. P., 1996,
BAAS 28, 822

\bibitem[1998]{platais98}
Platais I., Kozhurina-Platais V., van Leeuwen F., 1998, AJ 116, 2423

\bibitem[1992]{press}
Press W.H., Teukolsky S.A., Vetterling W.T., Flannery B.P., 1992, 
Numerical Recipes, Cambridge University Press

\bibitem[1998]{raboud}
Raboud D., Mermilliod J.C., 1998, A\&A 329, 101 

\bibitem[1997]{robichon97}
Robichon N., Arenou F., Turon C., Mermilliod J.C., 
Lebreton Y., 1997, ESA SP-402, 567

\bibitem[1994]{roeser}
R\"oser S., Bastian U., 1994, A\&A 285, 875

\bibitem[1992]{rosvick}
Rosvick J.M., Mermilliod J.C., Mayor M., 1992, A\&A 259, 720

\bibitem[1996]{tian}
Tian K.P., van Leeuwen F., Zhao J.L., Su C.G., 1996, 
A\&AS 118, 503

\bibitem[1992]{turon}
Turon C., Cr\'ez\'e M., \'Egret D. et al., 1992, The Hipparcos Input Catalogue, 
ESA SP-1136

\bibitem[1998]{act}
Urban S.E., Corbin T.E., Wycoff G.L., 1998, AJ 115, 2161

\bibitem[1999]{dezeeuw}
de Zeeuw T., Hoogerwerf R., de Bruijne J.H.J., Brown A.G.A.,
Blaauw A., 1999, AJ 117, 354
\end{thebibliography}
\end{document}